\documentclass[12pt]{article}

\input epsf

\begin{document}

\begin{center}
{\large \bf An $O(n^2\log^4 n \log \log n)$ Time Matrix Multiplication Algorithm}

{\it Yijie Han}

School of Science and Engineering\\
University of Missouri at Kansas City\\
Kansas City, MO 64110, USA\\
hanyij@umkc.edu

\vskip 0.2in
\parbox{4.5in}
{We show, for the input vectors $(a_0, a_1, ..., a_{n-1})$ and $(b_0, b_1, ..., b_{n-1})$, 
where $a_i$'s and $b_j$'s are real numbers, after $O(n\log^4 n)$ time preprocessing for each of them, the vector multiplication  
$(a_0, a_1, ..., a_{n-1})(b_0, b_1, ..., b_{n-1})^T $ can be computed in $O(\log^4 n \log \log n)$ time. 
This enables the
matrix multiplication for two $n\times n$ matrices 
to be computed in $O(n^2 \log^4 n\log \log n)$ time.

\noindent
Keywords: Algorithms, matrix multiplication, efficient algorithms.
}
\end{center}

\baselineskip=5.3mm

\section{Introduction}

Matrix multiplication is a fundamental computing task. Many researchers worked very hard to improve the
time complexity for matrix multiplication from the trivial $O(n^3)$ time to $n^{2.37286}$ time \cite{AW21}\cite{BCRL79}\cite{CW81}\cite{CW90}\cite{LeGall14}\cite{Pan78}\cite{CLRS09}\cite{Romani82}\cite{Schonhage}\cite{Stothers10}\cite{Strassen69}\cite{Williams12}. Notably Strassen first broke the $O(n^3)$ bound \cite{Strassen69}, Coppersmith and Winograd's $n^{2.376}$
bound \cite{CW90} lasted for more than twenty years, and the very recent bound of $n^{2.37286}$ achieved by Alman and Williams \cite{AW21}.

There are also works on multiplying rectangular matrices \cite{Coppersmith82}\cite{Coppersmith97}\cite{LeGall12}\cite{GU18}\cite{HP98}, with the recent result
of Le Gall and Urrutia \cite{GU18} showed that
an $n \times n^{0.31389}$ matrix can be multiplied with an $n^{0.31389} \times n$ matrix in $O(n^{2+\epsilon})$ time.

Matrix multiplication for integers can be used for multiplying Boolean matrices. For Boolean matrix multiplication and integer matrix multiplication there are no better algorithms other than those algorithms mentioned above. Some of the algorithms designed for multiplying Boolean matrices can be found 
in \cite{ADKF70}\cite{BW09}\cite{Chan15}\cite{Yu15}. A recent result of Yu \cite{Yu15} presented an $O(n^3 (\log \log n)^c /\log^4 n)$ time algorithm for Boolean matrix
multiplication. 

We show, for the input vectors $(a_0, a_1, ..., a_{n-1})$ and $(b_0, b_1, ..., b_{n-1})$, 
where $a_i$'s and $b_j$'s are real numbers, after $O(n\log^4 n)$ time preprocessing for each of them, the vector multiplication  
$(a_0, a_1, ..., a_{n-1})(b_0, b_1, ..., b_{n-1})^T$ can be computed in 
$O(\log^4 n\log \log n)$ time. 
Thus
matrix multiplication of two $n\times n$ matrices  
can be computed in $O(n^2 \log^4 n\log \log n)$ time. 

We assume that $n=2^k$ for a nonnegative integer $k$. Logarithms in this paper are of base 2. 

%We will do matrix multiplication for nonnegative input numbers. 
%That is we assume that all input numbers are nonnegative.
%To apply our algorithm for
%negative numbers we do  $(a_0, a_1, ..., a_{n-1})(b_0, b_1, ..., b_{n-1})^T=((a_0, a_1,..., a_{n-1})+N)((b_0, b_1, ..., b_{n-1})+N)^T-N\sum_{i=0}^{n-1}a_i-N\sum_{j=0}^{n-1}b_j-nN^2$, where $N$ is a large positive 
%number to bring every input number to be nonnegative. 

%I have written a C++ program to verify my algorithm. This C++ program is listed in the appendix of this paper. Because my %matrix multiplication algorithm is verified by this program and therefore it can be certain that my matrix multiplication %algorithm has no errors. 
 
\section{Version 1 of The Algorithm}

We use $\log n$ indeterminates $y_0, y_1, ..., y_{\log n-1}$. We will use $2\log n$ polynomials
$1-y_k$, $1+y_k$, $k=0, ...,\log n-1$. 
We will do the following one multiplication:\\

\noindent
$(1/2^{\log n})(\sum_{i=0}^{n-1}a_i\prod_{k=0}^{\log n-1}(1-y_k)^{i\#k}(1+y_k)^{\overline{i\#k}})$\\
$*(\sum_{j=0}^{n-1}b_j\prod_{k=0}^{\log n-1}
(1-y_k)^{j\#k}(1+y_k)^{\overline{j\#k}}) \; {\rm mod}\; (1-y_0^2)\; {\rm mod}\; (1-y_1^2) ... \; {\rm mod}\; (1-y_{\log n-1}^2) $ \hfill (1)\\

\noindent
where $i\#k$ ($j\#k$) is the $k$-th bit of $i$ ($j$) counting from the least significant bit starting at 0. $\overline{i\#k}=1-(i\#k)$ and $\overline{j\#k}=1-(j\#k)$. The $f_{i, j}(y_0, y_1, ..., y_{\log n-1})$ in each term $a_ib_jf_{i, j}(y_0, y_1, ..., y_{\log n-1})$ in (1) has degree 1 for $y_k$, $k=0, 1, ..., \log n-1$. Note that\\

\noindent
$f_{i, i}(y_0, y_1, ..., y_{\log n-1})=(1/2^{\log n})(\prod_{k_0 \in I_0} (1-y_{k_0})^2)(\prod_{k_1 \in I_1}(1+y_{k_1})^2)$\\
$\; {\rm mod}\; (1-y_0^2) \; {\rm mod}\; (1-y_1^2) ... \; {\rm mod}\; (1-y_{\log n-1}^2)  $\\
$=(1/2^{\log n})\prod_{k_0 \in I_0} (2-2y_{k_0}))\prod_{k_1 \in I_1}(2+2y_{k_1}))$ \hfill (2)\\

\noindent
where $I_0$ is the set indices of bits of $i$ which are 0's and $I_1$ is the set of indices of bits of $i$ which are 1's. For example $i=00101$ then $I_0=\{1, 3, 4\}$ and $I_1=\{ 0, 2\}$. \\ 

For $i\neq j$:\\

\noindent
$f_{i, j}(y_0, y_1, ..., y_{\log n-1})=(1/2^{\log n})( \prod_{k_{00} \in IJ_{00}} (1-y_{k_{00}})^2\prod_{k_{11} \in IJ_{11}}(1-y_{k_{11}})^2)$\\
$(\prod_{k_{01} \in IJ_{01}}(1-y_{k_{01}}^2))$\\
$(\prod_{k_{10}\in IJ_{10}}(1-y_{k_{10}}^2))  \; {\rm mod}\; (1-y_0^2)\; {\rm mod}\; (1-y_1^2) ... \; {\rm mod}\; (1-y_{\log n-1}^2) $ \hfill (3)\\

\noindent
Here $IJ_{00}$ is the set of indices of bits $k$ for $i$ and $j$ that $i_k=0$ and $j_k=0$,  
$IJ_{01}$ is the set of indices of bits $k$ for $i$ and $j$ such that $i_k=0$ and $j_k=1$,
$IJ_{10}$ is the set of indices of bits $k$ for $i$ and $j$ such that $i_k=1$ and $j_k=0$,
$IJ_{11}$ is the set of indices of bits $k$ for $i$ and $j$ such that $i_k=1$ and $j_k=1$.
For example $i=00101$ and $j=01001$, then $IJ_{00}=\{1, 4\}$, $IJ_{01}=\{3\}$, $IJ_{10}=\{2\}$, $IJ_{11}=\{0\}$.
 
Note that values for $IJ_{00}$ and $IJ_{11}$ need to be kept (set to nonzeros) and values for $IJ_{01}$ and $IJ_{10}$ need to be removed (set to 0's).

If $i\neq j$ then one of $IJ_{01}$ and $IJ_{10}$ is not empty and thus (3)
will become 0.

(2) will not be equal to 0. If we set $y_k=0$, for $k=0, 1, ..., \log n-1$, in (2) it will be equal to 1.

This algorithm is correct except it uses $\log n$ variables of degree 1 each.
These $\log n$ variables can be replaced by one variable of degree $2^{\log n}=n$.
This is the reason that the time complexity of this algorithm is not acceptable.

\section{Version 2 of The Algorithm}
This version 2 algorithm is incorrect and we show it for the explanation
of our algorithm. We will correct it in the version 3 of our algorithm.

In this version we use $\log \log n+1$ variables $x_0, x_1, ..., x_{\log \log n}$. We replace $y_i$ in our version 1 algorithm by $x_{\log \log n}\prod_{k=0}^{\log \log n-1}x_k^{i\#k}$. 

The version 2 algorithm will be:\\

\noindent
$(1/2^{\log n})(\sum_{i=0}^{n-1}a_i\prod_{k=0}^{\log n-1}((1-x_{\log \log n}\prod_{t=0}^{\log \log n-1} x_t^{k\#t})^{i\#k}(1+x_{\log \log n}\prod_{t=0}^{\log \log n-1}x_t^{k\#t})^{\overline{i\#k}}))$\\
$*(\sum_{j=0}^{n-1}b_j\prod_{k=0}^{\log n-1}
((1-x_{\log \log n}\prod_{t=0}^{\log \log n-1}x_t^{k\#t})^{j\#k}(1+x_{\log \log n}\prod_{t=0}^{\log \log n-1}x_t^{k\#t})^{\overline{j\#k}}))$\\
${\rm mod} \; (1-x_0^2)\; {\rm mod}\;  (1-x_1^2) ... \; {\rm mod}\; (1-x_{\log \log n}^2) $ \hfill (4)\\

This changes the number of variables to $\log \log n+1$ with degree $\log n$
for each variable. When replaced with one variable the degree becomes 
$(\log n)^{\log \log n+1}$. However, we used 
${\rm mod}\; (1-y_0^2)\; {\rm mod} \; (1-y_1^2) ... \; {\rm mod} \; (1-y_{\log n-1}^2) $
in our version 1 algorithm and this is replaced by  ${\rm mod}\; (1-x_0^2)\; {\rm mod} \; (1-x_1^2) ... \; {\rm mod}\;  (1-x_{\log \log n}^2) $ in this version 2 
algorithm. This changes the degree of each variable of $x_i$, $i=0, 1, ...,
\log \log n$, to 1. Thus when $x_i$, $i=0, 1, ...,
\log \log n$ are replaced with one variable the degree is $2\log n$. 

Our version 2 algorithm is incorrect. The problem is in the modulos. An example
will explain about this: $(1-x_0)^2(1-x_1)^2(1-x_0x_1)^2=0 \: {\rm mod} \; (1-x_0^2)
\; {\rm mod}\; (1-x_1^2)$. This should not be equal to 0 in (4) as in our version
1 algorithm $(1-y_1)^2(1-y_2)^2(1-y_3)^2 \neq 0 \; {\rm mod}\; (1-y_1^2)\; {\rm mod}
\; (1-y_2^2) \; {\rm mod} \; (1-y_3^2)$. This situation happens because
$(-x_0)(-x_1)(-x_0x_1)=-1 \; {\rm mod} \; (1-x_0^2) \; {\rm mod} \; (1-x_1^2)$.
If $\prod_k (-p_k)=-1 \; {\rm mod} \; (1-x_0^2) \; {\rm mod} \; (1-x_1^2) \; ...\; {\rm mod} (1-x_{\log \log n}^2)$ then $\prod_k (1-p_k) \; {\rm mod} \; (1-x_0^2) \; {\rm mod} \; (1-x_1^2) \; ...\; {\rm mod} (1-x_{\log \log n}^2) =0$.\\ 

\noindent
{\bf Condition of 0:} $\prod_{t=0}^{c} ((-1)^{k_t}y_{k_t})=-1  \;  {\rm mod}\; (1-x_0^2) \; {\rm mod}\; (1-x_1^2) ...
 \; {\rm mod}\; (1-x_{\log \log n}^2)$ is called the condition of 0 as it
forces $\prod_{t=0}^c (1+(-1)^{k_t}y_{k_t}) =0 \; {\rm mod} \; (1-x_0^2) \;
{\rm mod}\; (1-x_1^2) \; ... \; {\rm mod} \; (1-x_{\log \log n}^2)$.\\

For example: if $y_1y_2 =-1 \; {\rm mod}\; (1-x_0^2) \; {\rm mod}\; (1-x_1^2)\; ... \;
{\rm mod}\; (1-x_{\log \log n}^2)$, then $(1+y_1)(1+y_2)=1+y_1y_2+y_1(1+y_2/y_1)=1+y_1y_2+y_1(1+y_1y_2)=0\; {\rm mod}\; (1-x_0^2) \; {\rm mod}\; (1-x_1^2)\; ... \;
{\rm mod}\; (1-x_{\log \log n}^2)$.

It can be verified that if there is a factor $(1+p_0)(1+p_1)(1+p_2)\cdots (1+p_{k-1})$ in $a_ib_jf(i, x_0, x_1, ..., x_{\log \log n})f(j, x_0, x_1, ..., x_{\log\log  n})$,
then for anyone of $i=1, 2, ..., 2^k-1$, when $p_0^{i\#0}p_1^{i\#1}p_2^{i\#2}\cdots p_{k-1}^{i\#(k-1)}=-1 \; {\rm mod}\; (1-x_0^2) \; {\rm mod}\; (1-x_1^2)\; ... \;
{\rm mod}\; (1-x_{\log \log n}^2)$,  
where $p_k=\pm \prod_{i=0}^{c-1} x_{t_i}$ for a positive integer $c$ and $x_{t_i}$ being one of $x_0, x_1, ..., x_{\log n}$, 
then\\ 
$a_ib_jf(i, x_0, x_1, ..., x_{\log n})f(j, x_0, x_1, ..., x_{\log n})=0$, i.e. when  $p_0^{i\#0}p_1^{i\#1}p_2^{i\#2}\cdots p_{k-1}^{i\#(k-1)}=-1 \; {\rm mod}\; (1-x_0^2) \; {\rm mod}\; (1-x_1^2)\; ... \;
{\rm mod}\; (1-x_{\log \log n}^2)$, $(1+p_0)(1+p_1)(1+p_2)\cdots (1+p_{k-1})=0  \; {\rm mod}\; (1-x_0^2) \; {\rm mod}\; (1-x_1^2)\; ... \;
{\rm mod}\; (1-x_{\log \log n}^2)$.  

With $\log \log n+1$ variables $x_0, x_1, ...,
x_{\log \log n}$ there can be no more than $2*2^{\log \log n+1}-2=4\log n-2$ (factor 2 for positive and negative) different $f=(1 \pm  p_0^{i\#0}p_1^{i\#1}p_2^{i\#2}\cdots p_{k-1}^{i\#(k-1)})$'s . Because $f^2 =2f \; {\rm mod}\; (1-x_0^2) \; {\rm mod}\; (1-x_1^2)\; ...\; {\rm mod}\; (1-x_{\log \log n}^2)$ there are no more than
$4\log n-2$ different polynomials $(f(i, x_0, x_1, ..., x_{\log \log n}))^2$'s (here constant is not counted in as we view 
$(f(i, x_0, x_1, ..., x_{\log \log n}))^2$ and $c(f(i, x_0, x_1, ..., x_{\log \log n}))^2$ to be the same polynomial when $c$ is a nonzero constant). Because
we are summing $\sum_{k=0}^{n-1} a_ib_i(f(i, x_0, x_1, ..., x_{\log \log n}))^2$ and thus we need $f(i, x_0, x_1, ..., x_{\log \log n}) \neq f(j, x_0, x_1, ..., x_{\log \log n})$ when $i\neq j$. That is:
there need to be $n$ different $f$'s for $\sum_{i=0}^{n-1} a_ib_if_i$ and
thus $\log \log n+1$ variables are not enough. To have $n$ different $f$'s we need to have at least $\log n$ different $x_i$'s.
Thus we need to use $\log n$
variables. When these variables are replaced with one variable the degree is $n$. Thus this returns us to the version 1 of our algorithm using $\log n$ different
variables or using one variable of degree $n$.

We will correct this problem in our version 3 algorithm.

\section{Version 3 of The Algorithm}

We first present the version 3 algorithm with a problem and then show how to overcome this problem.

In this version we use $\log \log n+1$ variables $x_0, x_1, ..., x_{\log \log n}$. We replace $y_i$ in our version 1 algorithm by $x_{\log \log n}\prod_{k=0}^{\log \log n-1}x_k^{i\#k}$. 

The version 3 algorithm will be:\\

\noindent
$(\sum_{i=0}^{n-1}a_i\prod_{k=0}^{\log n-1}((2-y_k-1/y_k)^{i\#k}(2+y_k+1/y_k)^{\overline{i\#k}}))$\\
$(\sum_{j=0}^{n-1}b_j\prod_{k=0}^{\log n-1}((2-y_k-1/y_k)^{j\#k}(2+y_k+1/y_k)^{\overline{j\#k}})) \;$\\
$ {\rm mod}\; (1-x_0^2) \; {\rm mod}\; (1-x_1^2) ...
 \; {\rm mod}\; (1-x_{\log \log n}^2)$\\
$=(\sum_{i=0}^{n-1}a_if(i, x_0, x_1, ..., x_{\log \log n}))
(\sum_{j=0}^{n-1}b_jf(j, x_0, x_1, ..., x_{\log \log n})) \;$\\
$ {\rm mod}\; (1-x_0^2) \; {\rm mod}\; (1-x_1^2) ...
 \; {\rm mod}\; (1-x_{\log \log n}^2)$ \hfill (5)\\

Where $y_i$, $i=0, 1..., \log n-1$, is replaced by 
$x_{\log \log n}\prod_{k=0}^{\log \log n-1}x_k^{i\#k}$. (5) is a polynomial of
$2(\log \log n+1)$ variables ($x_i$'s and $1/x_i$'s) with each variable of degree $\log n$ ($x_k^0$ to $x_k^{\log n}$ or $1/x_i^0$ to $1/x_i^{\log n}$). Thus each of $f(i, x_0, x_1,..., x_{\log \log n})$ and
$f(j, x_0, x_1, ..., x_{\log \log n})$ is a polynomial of $2(\log \log n+1)$ variables of degree $\log n$ for each variable. However, we used  ${\rm mod}\; (1-x_0^2) \; {\rm mod}\; (1-x_1^2) ...
 \; {\rm mod}\; (1-x_{\log \log n}^2)$ and thus after modulos  each of $f(i, x_0, x_1,..., x_{\log \log n})$ and
$f(j, x_0, x_1, ..., x_{\log \log n})$ is a polynomials of $2(\log \log n+1)$ variables of degree 1 for each variable, and when replaced by one variable it is of degree
$2^{2(\log \log n+1)}$. Note that we cannot let $x_i=1/x_i$ and we cannot let $x_i/x_i=1$.

Note that for $i\neq j$ there is a bit $k$ such that $i\#k \neq j\#k$ and we have\\

\noindent
$(2-y_k-1/y_k)(2+y_k+1/y_k) \; {\rm mod}\; (1-x_0^2) \; {\rm mod}\; (1-x_1^2) ...
 \; {\rm mod}\; (1-x_{\log \log n}^2)=2(1-y_k/y_k)$ \hfill (6)\\

\noindent
Note that here we cannot let $y_k/y_k=1$ as this may conflict with ${\rm mod}\; (1-y_k^2)$ because $(y_k^2\; {\rm mod} \; (1-x_0^2) \; {\rm mod} \; (1-x_1^2) \; ... \; {\rm mod} \; (1-x_{\log \log n}^2))/y_k=1/y_k$ and if we let $y_k/y_k=1$ then $y_k^2/y_k=y_k$.
We need to have $(6)=0$ and this seems to be achievable by letting $y_k/y_k=1$. However, we will show that it is not that simple and this is the key problem
of this version which needs to be fixed.  

When $i\#k=j\#k=0$ we have:\\

\noindent
$(2-y_k-1/y_k)(2-y_k-1/y_k) \; {\rm mod}\; (1-x_0^2) \; {\rm mod}\; (1-x_1^2) ...
 \; {\rm mod}\; (1-x_{\log \log n}^2)$\\
$=(2-y_k-1/y_k)^2 \; {\rm mod}\; (1-x_0^2) \; {\rm mod}\; (1-x_1^2) ...
 \; {\rm mod}\; (1-x_{\log \log n}^2)$\\
$=(6+2y_k/y_k-4y_k-4/y_k)$ \hfill (7)\\

\noindent
And when $i\#k=j\#k=1$ we have:\\

\noindent
$(2+y_k+1/y_k)(2+y_k+1/y_k) \; {\rm mod}\; (1-x_0^2) \; {\rm mod}\; (1-x_1^2) ...
 \; {\rm mod}\; (1-x_{\log \log n}^2)$\\
$=(2+y_k+1/y_k)^2 \; {\rm mod}\; (1-x_0^2) \; {\rm mod}\; (1-x_1^2) ...
 \; {\rm mod}\; (1-x_{\log \log n}^2)$\\
$=(6+2y_k/y_k+4y_k+4/y_k)$ \hfill (8)\\

Note here we can do $x_i - x_i=0$, $x_i^2=1 \; {\rm mod}\; (1-x_i^2)$,  $(1/(x_i))^2=1 \; {\rm mod}\; (1-x_i^2)$. But we cannot do $x_i=1/x_i \; {\rm mod} \; (1-x_i^2)$.  
We cannot let $x_i - 1/x_i$ equal to 0. These rules we specified here keep our
operations commutative and associative. That is we
have $f_1f_2f_3=(f_1f_2)f_3=f_1(f_2f_3)=f_1f_3f_2$.\\

\noindent
{\bf Lemma 1:}
$(f(i, x_0, x_1, ..., x_{\log \log n}))^2 {\rm mod}\; (1-x_0^2) \; {\rm mod}\; (1-x_1^2) ...
 \; {\rm mod}\; (1-x_{\log \log n}^2) \neq 0$ $\framebox{}$ \\ 

Only $(2+p+1/p)(2-p-1/p)$ can result in $2(1-p/p)$, where $p$ is a monomial of
$x_0, x_1, ..., x_{\log \log n}$ with degree $\leq 1$ for each $x_i$, which will result in 0 if we let $p/p=1$ (we will show there is a problem here and how to overcome it).
In $(f(i, x_0, x_1, ..., x_{\log \log n}))^2$ there is no such factor. When there is $(2+p+1/p)$ (there is $(2-p-1/p)$) in $(f(i, x_0, x_1, ..., x_{\log \log n}))^2$
there is no $(2-p-1/p)$ (there is no $(2+p+1/p)$) in  $(f(i, x_0, x_1, ..., x_{\log \log n}))^2$ thus $(f(i, x_0, x_1, ..., x_{\log \log n}))^2 {\rm mod}\; (1-x_0^2) \; {\rm mod}\; (1-x_1^2) ...
 \; {\rm mod}\; (1-x_{\log \log n}^2) \neq 0$ and Lemma 1 holds.

Note that the Condition 0 in our version 2 algorithm has been
changed here as we cannot have
$\prod_{t=0}^c ((-y_{k_t}-1/y_{k_t})/2)=-1 \; {\rm mod} \; (1-x_0^2) \; {\rm mod} \; (1-x_1^2) \; ...\; {\rm mod} \; (1-x_{\log \log n}^2)$ in order to have $\prod_{t=0}^c (1-y_{k_t}/2-1/(2y_{k_t})) \; {\rm mod} \; (1-x_0^2) \; {\rm mod} \; (1-x_1^2) \; ...\; {\rm mod}\; (1-x_{\log \log n}^2) =0$.

Note that $((y_i+1/y_i)/2)^2=(1+y_i/y_i)/2 \; {\rm mod} \; (1-y_i^2)$. Let $p_i=y_i+1/y_i$, we have\\

\noindent
$((1+y_i/y_i)/2)(1-p_i)(1-p_j)$\\
$=p_i^2(1+p_ip_j-(p_i+p_j))$\\
$=p_i^2(1+p_ip_j)-p_i^2(p_i+p_j)$\\
$=p_i^2(1+p_ip_j)-p_i(p_i^2+p_ip_j)$\\
$=p_i^2(1+p_ip_j)-p_i(1/2+y_i/(2y_i)+p_ip_j)$\\
$=p_i^2(1+p_ip_j)-p_i(1/2+p_ip_j/2)-p_i(y_i/y_i+p_ip_j)/2$ \hfill (9)\\

We know that if $-p_i=-1$ or $-p_j=-1$ then $(1-p_i)(1-p_j)=0$. Now if $(-p_i)(-p_j)=p_ip_j=-1$ then\\

\noindent
$(9)$\\
$=-p_i(y_i/y_i-1)/2$\\
$=-(y_i+1/y_i)(y_i/y_i-1)/2$\\
$=-(1/y_i-y_i+y_i-1/y_i)/2=0$\\

That is: if $p_ip_j=-1$ then $((1+y_i/y_i)/2)(1-p_i)(1-p_j)=0$, thus either $(1-p_i)(1-p_j)=c(1-y_i/y_i)$ for a constant $c$ or
$(1-p_i)(1-p_j)=0$. Since $(1-p_i)(1-p_j) \neq c(1-y_i/y_i)$ thus we conclude that $(1-p_i)(1-p_j)=0$. That is $-p_i=-1$, $-p_j=-1$
and $(-p_i)(-p_j)=p_ip_j=-1$ will all result in $(1-p_i)(1-p_j)=0 \; {\rm mod} \; (1-x_0^2) \; {\rm mod}\; (1-x_1^2) \; ...\; {\rm mod}\;
(1-x_{\log \log n}^2)$. Thus the Condition of 0 that holds for $y_i$'s holds for $p_i$'s as well. The difference is that 
$\prod_{t=0}^{c} ((-1)^{k_t}y_{k_t})=-1  \;  {\rm mod}\; (1-x_0^2) \; {\rm mod}\; (1-x_1^2) ...
 \; {\rm mod}\; (1-x_{\log \log n}^2)$ can hold but  $\prod_{t=0}^{c} ((-1)^{k_t}p_{k_t})=-1  \;  {\rm mod}\; (1-x_0^2) \; {\rm mod}\; (1-x_1^2) ...
 \; {\rm mod}\; (1-x_{\log \log n}^2)$ will never hold. Thus $(f(i, x_0, x_1, ..., x_{\log \log n}))^2$ will not be equal to 0.\\

\noindent
{\bf Example:}

\noindent
$(1+x_0)^2(1-x_1)^2(1+x_0x_1)^2=0 \; {\rm mod} \; (1-x_0^2) \; {\rm mod} \; (1-x_1^2)$ \hfill (10)\\

\noindent
but\\

\noindent
$(2+x_0+1/x_0)^2(2-x_1-1/x_1)^2(2+x_0x_1+1/x_0x_1)^2 \; {\rm mod} \; (1-x_0^2) \; {\rm mod} \; (1-x_1^2)$\\
$=(6+2x_0/x_0+4x_0+4/x_0)(6+2x_1/x_1-4x_1-4/x_1)(6+2x_0x_1/(x_0x_1)+4x_0x_1+4/(x_0x_1))$\\
$ {\rm mod} \; (1-x_0^2) \; {\rm mod} \; (1-x_1^2)$\\
$=6*(36+12x_1/x_1+12x_0/x_0+4x_0x_1/(x_0x_1))-64(1+x_0/x_0+x_1/x_1+x_0x_1/(x_0x_1))$\\
$-16*6(x_0x_1+1/(x_0x_1)+x_0/x_1+x_1/x_0)+144x_0x_1+16/(x_0x_1)+48(x_0/x_1+x_1/x_0)$\\
$-(48(x_0/(x_0x_1)+x_0x_1/x_0)+16(x_1+1/x_1))+96(x_0/(x_0x_1)+1/x_1)+32(x_0x_1/x_0+x_1)$\\
$-32(x_0x_1+x_0/x_1+x_1/x_0+1/(x_0x_1))+144/(x_0x_1)+48(x_0/x_1+x_1/x_0)+16x_0x_1$\\
$+48(x_1/(x_0x_1)+x_0x_1/x_1)+16(x_0+1/x_0)-(96(x_0+x_0x_1/x_1)+32(1/x_0+x_1/(x_0x_1)))$\\
$+6*(24(x_0+1/x_0)+8(x_0x_1/x_1+x_1/(x_0x_1)))-(96(1/x_0+x_1/(x_0x_1))+32(x_0+x_0x_1/x_1))$\\
$+72x_0x_1/(x_0x_1)+24(x_0/x_0+x_1/x_1)+8-64(1+x_0/x_0+x_1/x_1+x_0x_1/(x_0x_1))$\\
$-6*(24*(x_1+1/x_1)+8(x_0x_1/x_0+x_0/(x_0x_1)))+96(x_1+x_0x_1/x_0)+32(1/x_1+x_0/(x_0x_1))$\\
$=96-32x_1/x_1-32x_0/x_0-32x_0x_1/(x_0x_1)$\\
$+32x_0x_1+32/(x_0x_1)-32x_0/x_1-32x_1/x_0$\\
$+32x_0+32/x_0-32x_1x_0/x_1-32x_1/(x_0x_1)$\\
$-32x_1-32/x_1+32x_0x_1/x_0+32x_0/(x_0x_1)$ \hfill (11)\\
$=32x_0x_1+32/(x_0x_1)-32x_0/x_1-32x_1/x_0$ \hfill (12)\\

As we see here that $(10)=0$ but $(11)\neq 0$ and $(12) \neq 0$. From (11) to (12) we let $x_i/x_i=1$.

The problem with (5) as we mentioned earlier is that $f(i, x_0, x_1, ..., x_{\log \log n})f(j, x_0, x_1, ..., x_{\log \log n})$ may not equal to 0
when $i \neq j$. As an example if we multiply $(1-x_0/x_0)$ in (6) to (11) we get\\

\noindent
$96-32x_1/x_1-32x_0/x_0-32x_0x_1/(x_0x_1)$\\
$+32x_0x_1+32/(x_0x_1)-32x_0/x_1-32x_1/x_0$\\
$+32x_0+32/x_0-32x_1x_0/x_1-32x_1/(x_0x_1)$\\
$-32x_1-32/x_1+32x_0x_1/x_0+32x_0/(x_0x_1)$\\
$-96x_0/x_0+32x_0x_1/(x_0x_1)+32+32x_1/x_1$\\
$-32x_1/x_0-32x_0/x_1+32/(x_0x_1)+32x_0x_1$\\
$-32/x_0-32x_0+32x_1/(x_0x_1)+32x_0x_1/x_1$\\
$+32x_0x_1/x_0+32x_0/(x_0x_1)-32x_1-32/x_1$\\
$=64x_0x_1+64/(x_0x_1)-64x_0/x_1-64x_1/x_0$\\
$\neq 0$ \hfill (13)\\

That is we cannot use (6) to force  $f(i, x_0, x_1, ..., x_{\log \log n})f(j, x_0, x_1, ..., x_{\log \log n})$ equal to 0 when $i \neq j$.

Now we will overcome this problem. We change (5) to:\\

\noindent
$(\sum_{i=0}^{n-1}a_i\prod_{k=0}^{\log n-1}(((1-{\rm \bf i}y_k/y_k)(2-y_k^2-1/y_k^2))^{i\#k}((1+{\rm \bf i}y_k/y_k)(2+y_k^2+1/y_k^2))^{\overline{i\#k}}))$\\
$(\sum_{j=0}^{n-1}b_j\prod_{k=0}^{\log n-1}(((1-{\rm \bf i}y_k/y_k)(2-y_k^2-1/y_k^2))^{j\#k}((1+{\rm \bf i}y_k/y_k)(2+y_k^2+1/y_k^2))^{\overline{j\#k}})) \;$\\
$ {\rm mod}\; (1-x_0^4) \; {\rm mod}\; (1-x_1^4) ...
 \; {\rm mod}\; (1-x_{\log \log n}^4)$\\
$=(\sum_{i=0}^{n-1}a_if(i, x_0, x_1, ..., x_{\log \log n}))
(\sum_{j=0}^{n-1}b_jf(j, x_0, x_1, ..., x_{\log \log n})) \;$\\
$ {\rm mod}\; (1-x_0^4) \; {\rm mod}\; (1-x_1^4) ...
 \; {\rm mod}\; (1-x_{\log \log n}^4)$ \hfill (14)\\

\noindent
here ${\rm \bf i} = \sqrt{-1}$. The definition of $y_k$ has not been changed and it remains to be $y_k=x_{\log \log n}\prod_{i=0}^{\log \log n-1}x_i^{k\#i}$. We changed $y_k$ to $y_k^2$ and changed ${\rm mod} \; (1-x_i^2)$ to ${\rm mod}\; (1-x_i^4)$.

In (14) we basically used $2(\log \log n+1)$ variables $x_0, 1/x_0, x_1, 1/x_1, ..., x_{\log \log n}, 1/x_{\log \log n}$ with degree $4\log n$ (range from $0$
to $4\log n$) for each variable. Because we used 
${\rm mod}\; (1-x_0^4) \; {\rm mod} \; (1-x_1^4) \; ... \; {\rm mod} \; (1-x_{\log \log n}^4)$ and therefore each variable has degree 3 (0 to 3). When converted
to one variable it is of degree $4^{2(\log \log n+1)}=16\log^4 n$ (each $x_i$ ($1/x_i$) can take $x_i^t$ ($1/x_i^t$) for $t=0, 1, 2, 3$).    

(6) now becomes:\\

\noindent
$(1-{\rm \bf i}y_k/y_k)(2-y_k^2-1/y_k^2)(1+{\rm \bf i}y_k/y_k)(2+y_k^2+1/y_k^2)$\\ 
$\; {\rm mod}\; (1-x_0^4) \; {\rm mod}\; (1-x_1^4) ...
 \; {\rm mod}\; (1-x_{\log \log n}^4)$\\
$=2(1+y_k^2/y_k^2)(1-y_k^2/y_k^2) \; {\rm mod}\; (1-x_0^4) \; {\rm mod}\; (1-x_1^4) ...
 \; {\rm mod}\; (1-x_{\log \log n}^4)$\\
$=2(1-y_k^4/y_k^4) \; {\rm mod}\; (1-x_0^4) \; {\rm mod}\; (1-x_1^4) ...
 \; {\rm mod}\; (1-x_{\log \log n}^4)$\\
$=0$ \hfill (15)\\

\noindent
Thus now (15) can force $f(i, x_0, x_1, ..., x_{\log \log n})f(j, x_0, x..., x_{\log \log n})$ to become 0 when $i \neq j$.  

When $i\#k=j\#k=0$ (7) becomes\\

\noindent
$(1-{\rm \bf i}y_k/y_k)(2-y_k^2-1/y_k^2)(1-{\rm \bf i}y_k/y_k)(2-y_k^2-1/y_k^2) \; {\rm mod}\; (1-x_0^4) \; {\rm mod}\; (1-x_1^4) ...
 \; {\rm mod}\; (1-x_{\log \log n}^4)$\\
$=((1-{\rm \bf i}y_k/y_k)(2-y_k^2-1/y_k^2))^2 \; {\rm mod}\; (1-x_0^4) \; {\rm mod}\; (1-x_1^4) ...
 \; {\rm mod}\; (1-x_{\log \log n}^4)$\\
$=(1-y_k^2/y_k^2-2{\rm \bf i}y_k/y_k)(6+2y_k^2/y_k^2-4y_k^2-4/y_k^2)\; {\rm mod}\; (1-x_0^4) \; {\rm mod}\; (1-x_1^4) ...
 \; {\rm mod}\; (1-x_{\log \log n}^4)$\\
$=4(1-y_k^2/y_k^2)-{\rm \bf i}(2y_k/y_k)(6+2y_k^2/y_k^2-4y_k^2-4/y_k^2)\; {\rm mod}\; (1-x_0^4) \; {\rm mod}\; (1-x_1^4) ...
 \; {\rm mod}\; (1-x_{\log \log n}^4)$ \hfill (16)\\

\noindent
And when $i\#k=j\#k=1$ (8) becomes\\

\noindent
$(1+{\rm \bf i}y_k/y_k)(2+y_k^2+1/y_k^2)(1+{\rm \bf i}y_k/y_k)(2+y_k^2+1/y_k^2) \; {\rm mod}\; (1-x_0^4) \; {\rm mod}\; (1-x_1^4) ...
 \; {\rm mod}\; (1-x_{\log \log n}^4)$\\
$=((1+{\rm \bf i}y_k/y_k)(2+y_k^2+1/y_k^2))^2 \; {\rm mod}\; (1-x_0^4) \; {\rm mod}\; (1-x_1^4) ...
 \; {\rm mod}\; (1-x_{\log \log n}^4)$\\
$=(1-y^2/y^2+2{\rm \bf i}y_k/y_k)(6+2y_k^2/y_k^2+4y_k^2+4/y_k^2)  \; {\rm mod}\; (1-x_0^4) \; {\rm mod}\; (1-x_1^4) ...
 \; {\rm mod}\; (1-x_{\log \log n}^4)$\\
$=4(1-y_k^2/y_k^2)+{\rm \bf i}(2y_k/y_k)(6+2y_k^2/y_k^2+4y_k^2+4/y_k^2)  \; {\rm mod}\; (1-x_0^4) \; {\rm mod}\; (1-x_1^4) ...
 \; {\rm mod}\; (1-x_{\log \log n}^4)$ \hfill (17)\\

$(f(i, x_0, x_1, ..., x_{\log \log n}))^2$ will not be equal to 0 because if we let
$y_k^2/y_k^2=1$ for $k=0, ..., \log n-1$, which is compatible to ${\rm mod}\; (1-x_0^4) \; {\rm mod}\; (1-x_11^4) \;...\; {\rm mod}\; (1-x_{\log \log n}^4)$, then (16) becomes\\

\noindent
$-{\rm \bf i}(2y_k/y_k)(8-4y_k^2-4/y_k^2)\; {\rm mod}\; (1-x_0^4) \; {\rm mod}\; (1-x_1^4) ...
 \; {\rm mod}\; (1-x_{\log \log n}^4)$ \hfill (18)\\

\noindent
And (17) becomes\\

\noindent
${\rm \bf i}(2y_k/y_k)(8+4y_k^2+4/y_k^2)  \; {\rm mod}\; (1-x_0^4) \; {\rm mod}\; (1-x_1^4) ...
 \; {\rm mod}\; (1-x_{\log \log n}^4)$ \hfill (19)\\

Thus\\

\noindent
$(f(i, x_0, x_1, ..., x_{\log \log n}))^2 \; {\rm mod}\; (1-x_0^4) \; {\rm mod}\; (1-x_1^4) ...
 \; {\rm mod}\; (1-x_{\log \log n}^4)$\\
$=\prod_{k=0}^{\log n-1}((-{\rm \bf i}(2y_k/y_k)(8-4y_k^2-4/y_k^2))^{i\#k}({\rm \bf i}(2y_k/y_k)(8+4y_k^2+4/y_k^2))^{\overline{i\#k}})$\\
$ {\rm mod}\; (1-x_0^4) \; {\rm mod}\; (1-x_1^4) ...
 \; {\rm mod}\; (1-x_{\log \log n}^4)$ \hfill (20)\\

(20) is not equal to 0 as we have studied about the correctness of Lemma 1 and demonstrated in (11) and (12) (the factors $y_k/y_k$'s cannot make it to become 0). 

After\\

\noindent 
$(f(i, x_0, x_1, ..., x_{\log \log n}))^2$\\
${\rm mod}\; (1-x_0^4) \; {\rm mod}\; (1-x_1^4) ...
 \; {\rm mod}\; (1-x_{\log \log n}^4)$  \hfill (21)\\

\noindent
has been computed, we then set $x_k=x^{4^{2k+2}}$, set
$1/x_k=x^{4^{2k+1}}$ and set ${\rm \bf i}=x^{4^{2\log \log n+4}}$. This will turn (21) to a nonzero polynomial $c_i(x)$ with only one variable $x$ of
degree $4^{2(2\log \log n+4)}=2^{16}\log^8 n$. We then set $x=40^{\log n}=n^{\log 40}$ and $c_i(x)$ will not equal to 0 as if we let $y_k=1$, ${\rm \bf i}=1$ and
change negative and minus to positive and plus, 
then $4(1-y_2^2/y_2^2)\pm 2{\rm \bf i}y_2/y_2)(6+2y_2^2/y_2^2\pm (4y_2^2+4/y_2^2))$ will become  $4(1+1)+ 2(6+2+ 4+4)=40$. Note also that although $c_i(x)$ is a degree
$2^{16}\log^8 n$ polynomial, the evaluation of $c_i(c)$ takes $O(\log^4 n)$ time
because ${\rm \bf i}=x^{4^{2\log \log n+4}}$ and $c_i(x)=c_1(x)+{\rm \bf i}c_2(x)$
before ${\rm \bf i}$ is replaced by $x^{4^{2\log \log n+4}}$, where $c_1(x)$
and $c_2(x)$ both have degree $4^{2\log \log n+2}=16\log^4 n$.

We multiply $a_i$ with $1/c_i(n^{\log 40})$ before we compute (14). After we computed (14) we 
set $x_i=n^{(\log 40)4^{2i+2}}$, set $1/x_i=n^{(\log 40)4^{2i+1}}$ and set ${\rm \bf i}=n^{(\log 40)4^{\log \log n+4}}$
and this will give us the value of $(a_0, a_1, ..., a_{n-1})(b_0, b_1, ..., b_{n-1})^T$.\\

\noindent 
{\bf Lemma 2:} After $O(n\log^4 n)$ time preprocessing for vectors $V_L=(a_0, a_1, ..., a_{n-1})$ and
$V_R=(b_0, b_1, ..., b_{n-1})$ with real numbers, $V_LV_R^T$ can be computed in $O(\log^4 n\log \log n)$ time. $\framebox{}$\\

\noindent
{\bf Note:} Because each $a_i$ ($b_j$) needs to be multiplied with a polynomial of $2(\log \log n+1)$ variables with degree 4 for each variable,
this is 
for a single variable of degree  $O(4^{2(\log \log n+1)})=O(n^4)$. The multiplication takes 
$O(\log^4 n\log \log n)$ time as we can multiply two degree $O(\log^4 n)$ polynomials in $O(\log^4 n\log \log n)$ time using Fast Fourier Transform \cite{CLRS09}.\\

\noindent
{\bf Main Theorem:} Two $n \times n$ matrices with real numbers can be multiplied in $O(n^2\log^4 n\log \log n)$ time. $\framebox{}$\\

\noindent
{\bf Note:} As per Lemma 2 the preprocessing takes $O(n^2 \log^4 n)$ time. The $n^2$ multiplications
of two degree $O(\log^4 n)$ polynomials takes $O(\log^4 n\log \log n)$ time using Fast Fourier Transform \cite{CLRS09}
for each multiplication.\\

\section{Conclusion}
Our method provides a new approach to solving the matrix multiplication
problem. Our vector multiplication took $O(\log^2 \log \log n)$ time. This could be the optimal time complexity but the algorithm had the problem of unable to force $f(i, x_0, x_1, ..., x_{\log \log n})f(j, x_0, x_1,...,
x_{\log \log n})$ to 0 when $i\neq j$. We fixed this problem with an additional factor $\log^2 n$ of time complexity to bring the vector mutliplication
time complexity to $O(\log^4 n \log \log n)$.

In a sense our vector and matrix multiplication results are obtained from 0's as if we let $1/x_i=x_i \; {\rm mod} \; (1-x_i^2)$ then (11) and (12) are equal to 0's. 

I feel that I have achieved something impossible. 

We are investigating whether $O(n^2\log n)$ or $O(n^2)$ time is reachable for matrix multiplication. We are now getting pretty close to this goal. 

In another paper \cite{Han} I proved $P=NP$. This proof heavily depends on the vector and matrix multiplication algorithm described in this paper. 

%\section{}
%\label{}





%\bibliographystyle{alpha}

\end{document}